\documentclass[%
 reprint,
 amsmath,amssymb,
 aps,
]{revtex4-2}
\usepackage{color,xcolor}
\usepackage{soul}
\usepackage{graphicx}
\usepackage{dcolumn}
\usepackage{bm}
\usepackage[table]{xcolor}
\usepackage{array}
\usepackage{booktabs}
\usepackage{multirow}
\usepackage{siunitx}
\usepackage{graphicx}
\usepackage{graphicx}
\usepackage{subcaption}
\usepackage{graphicx}
\usepackage{overpic}
\usepackage{caption}
\usepackage[colorlinks=true, citecolor=blue, linkcolor=blue, urlcolor=blue]{hyperref}
\usepackage[colorlinks=true, citecolor=blue, linkcolor=blue, urlcolor=blue]{hyperref}

\begin{document}

\preprint{APS/123-QED}

\title{Lattice-Expansion-Driven Stabilization of Helical Magnetic Order in Ru-Doped MnP}

\author{Xin-Wei Wu}
\author{Denglu Hou}
\author{Li Ma}%
\author{Congmian Zhen}
\author{Dewei Zhao}
 \author{Guoke Li}
  \email{Contact author: liguoke@126.com}

\affiliation{%
Hebei Advanced Thin Films Laboratory, College of Physics, Hebei Normal University, Shijiazhuang, 050024, Hebei, China}%

\date{\today}

\begin{abstract}
The practical utilization of MnP in chiral spintronic devices is fundamentally constrained by its low helical ordering temperature ($T_{\mathrm{\textit{S}}}$). Here, we demonstrate that Ru substitution in Mn$_{1-\textit{x}}$Ru$_\textit{x}$P single crystals drives a highly anisotropic lattice expansion, where in the $\textit{b}$-axis elongation is one-quarter that of the $a\textit{}$- and $\textit{c}$-axis ($\sim$ 0.04 \AA). This structural distortion profoundly stabilizes the helical ground state, elevating $T_{\mathrm{\textit{S}}}$ from 51 K to 215 K and the critical field along [010] direction at 5 K from 2.3 to 30.0 kOe, while suppressing the Curie temperature ($T_{\mathrm{\textit{C}}}$) from 291 K to 215 K. Synthesizing these results with reported data on Mo- and W-doped analogues reveals that $T_{\mathrm{\textit{S}}}$ and $T_{\mathrm{\textit{C}}}$ are governed primarily by the $\textit{b}$-axis parameter, exhibiting universal linear scaling relationships ($\frac{dT_\textit{S}}{d\textit{b}}=1.59 \times 10^4\ \text{K·Å}^{-1} $ , $\frac{dT_\textit{C}}{d\textit{b}}=0.69 \times 10^4\ \text{K·Å}^{-1}$ ) far greater than those associated with the $\textit{a}$- or $\textit{c}$-axes. First-principles calculations reveal that the lattice expansion selectively attenuates ferromagnetic coupling while preserving antiferromagnetic interactions between nearest-neighbor Mn atoms, thereby enhancing magnetic frustration and stabilizing helimagnetism. These findings establish chemical pressure via directed $\textit{b}$-axis engineering as a robust, generalizable paradigm for stabilizing helimagnetism in MnP.
\end{abstract}

\maketitle


\section{INTRODUCTION}
MnP crystallizes in a distorted NiAs-type structure, featuring a distorted triangular Mn sublattice within the bc-plane and zigzag Mn chains along the a-axis\cite{1,2}. As depicted in Fig.~1(a), this distortion engenders four distinct nearest-neighbor Mn--Mn distances\cite{3}, fostering a delicate competition between ferromagnetic and antiferromagnetic exchange interactions that underpins its remarkably rich magnetic phase diagram\cite{4,5,6,7}. In zero field,  MnP stabilizes into a double-helical antiferromagnetic order that transitions to ferromagnetism at $T_{\text{S}} = 50$ K before becoming paramagnetic at $T_{\text{C}} = 291$ K\cite{8,9,10}.Under applied magnetic fields, this fragile helical order evolves into intricate conical and fan phases, as well as topological skyrmion and soliton lattices\cite{11,12,13}. These exotic textures host a suite of emergent phenomena, such as the topological Hall effect\cite{14}, giant magneto-resistance\cite{15,16}, and current- and field-driven coherent control of helical chirality\cite{17,18}, positioning MnP as a prime candidate for next-generation spintronics. However, the low helical ordering temperature remains a fundamental bottleneck for practical device integration, necessitating robust strategies to expand the thermal stability window of chiral phase.

The magnetic transition temperatures of MnP exhibit extraordinary sensitivity to lattice distortions, suggesting strain engineering as a promising route to tune its helical order\cite{19,20,21}.Indeed, uniaxial pressure applied along the $c$-axis enhances $T_{\text{S}}$ while suppressing $T_{\text{C}}$, whereas pressure along the $a$- or $b$-axis produces opposite effects\cite{22}.  This tunability is further corroborated in thin films, where biaxial strain can double $T_{\text{S}}$ to $\sim 110$ K with minimal impact on $T_{\text{C}}$\cite{23,25,25}. In addition, hydrostatic pressure drives lattice contraction that initially reduces both $T_{\text{S}}$ and $T_{\text{C}}$, before inducing a transition to a new $b$ axis helical phase at $\sim 2.0$ GPa and the emergence of the superconductivity at $7.5$ GPa\cite{6,26}. Furthermore, while substituting Mn with Cr, Fe, Co, and V yields non-monotonic variations, while much larger dopants such as Mo and W sharply enhance $T_{\text{S}}$ and suppress $T_{\text{C}}$\cite{27,28,29,30}, suggesting that lattice expansion preferentially stabilizes the helical phase. Despite these empirical successes, the microscopic mechanism linking specific lattice distortions to the rebalancing of ferromagnetic and antiferromagnetic exchange interactions remains elusive. Crucially, it is unclear to what extent lattice parameters---distinct from electronic doping effects---govern the stabilization of the helical order across diverse chemical strategies

\begin{figure*}[htbp!]
    \centering
    \includegraphics[width=0.85\textwidth]{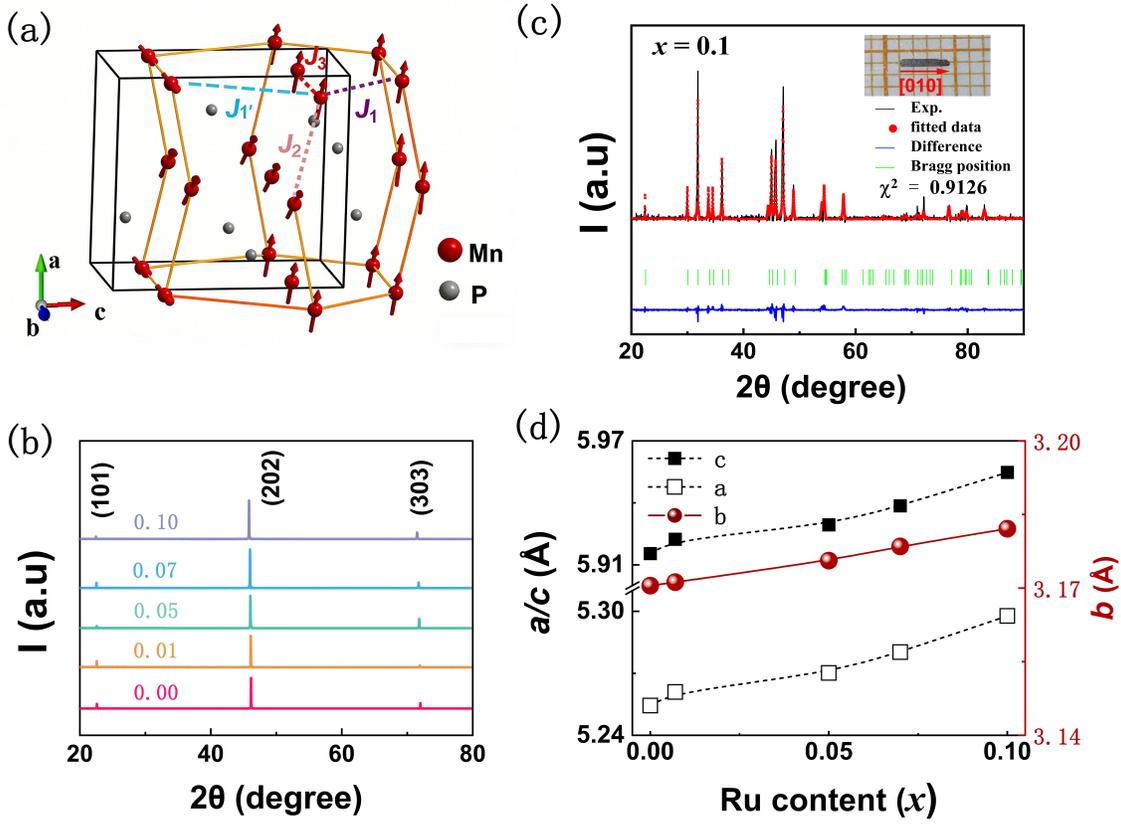}
    \vspace{0pt} 
    \caption{Structural characterization of Mn$_{1-x}$Ru$_x$P single crystals. (a) The crystal structure of MnP (space group \textit{Pnma}; $a = 5.260$\,Å; $b = 3.174$\,Å; $c = 5.919$\,Å), where the exchange parameters are depicted on a distorted NiAs-type lattice. (b) XRD patterns collected from the side facets of needle-like Mn$_{1-x}$Ru$_x$P single crystals for various Ru content $x$. (c) Representative Rietveld refinement profile of the XRD pattern for powder obtained by grinding single crystals. The inset shows an optical image of a typical Mn$_{1-x}$Ru$_x$P ($x = 0.00$, 0.01, 0.05, 0.07, 0.10) single crystal. (d) Evolution of the lattice parameters $a$, $b$, and $c$ with increasing $x$ from 0.00 to 0.10.}
    \vspace{0pt} 
    \label{Fig. 1}
\end{figure*}

 In this work, we demonstrate that increasing the Ru content in Mn$_{1-x}$Ru$_x$P ($0 \leq x \leq 0.10$) single crystals induces an anisotropic lattice expansion, driving a remarkable convergence of $T_{\text{S}}$ and $T_{\text{C}}$ at $215$ K and quadrupling the thermal stability window of the helical phase. Strikingly, we uncover a universal evolution of magnetic transition temperatures governed by the expansion of the $b$-axis lattice parameter, independent of the dopant species. First-principles calculations show that lattice expansion selectively suppresses ferromagnetic exchange while preserving antiferromagnetic coupling, thus amplifying magnetic frustration to intrinsically stabilize helimagnetism.

\section{Experimental}

Single crystals of Mn$_{1-x}$Ru$_x$P ($x = 0.00$, $0.01$, $0.05$, $0.07$, $0.10$) were grown using the Sn-flux method. High-purity Mn, Ru, P, and Sn ($> 99.99\%$) were loaded into an alumina crucible with a molar ratio of $(1-x):x:1:25$ ($x = 0.00$, $0.01$, $0.05$, $0.07$, $0.10$). The crucible was sealed in a quartz tube under $10^{-4}$~Pa, heated to $650\,^\circ$C for $12$~h to initiate reaction, then ramped to $1100\,^\circ$C and held for $7$~h for complete dissolution. The melt was slowly cooled at $4\,^\circ$C/h to $600\,^\circ$C, followed by centrifugation to remove Sn flux, yielding needle-like black single crystals (inset of Fig.1 (b)). The Ru doping levels, as quantified by energy-dispersive X-ray spectroscopy (EDS; Bruker XFlash 7/60), increased from $0.00$ to $0.10$. The crystal structure of the samples was characterized by X-ray diffraction (XRD, PANalytical) with Cu $K\alpha$ irradiation. Magnetic properties were measured using a Physical Property Measurement System (PPMS-9, Quantum Design).

To investigate the effect of lattice expansion, first-principles calculations were performed using VASP code. To accommodate the helical phase, a $1 \times 1 \times 9$ supercell based on the experimental lattice parameters of Mn$_{1-x}$Ru$_x$P was constructed without explicitly including Ru atoms. A $6 \times 6 \times 1$ Monkhorst–Pack $k$-point mesh and a $400$\,eV plane-wave cutoff was employed, with electronic self-consistency converged to $10^{-6}$\,eV per cell. The interchain exchange parameters $J_1$ and $J_1'$, as well as $J_2$ along the Mn zigzag chain were extracted using a $1 \times 1 \times 2$ MnP supercell. Four spin configurations—ferromagnetic, interchain ferromagnetic, cross-antiferromagnetic, and intrachain antiferromagnetic—were constructed, their total energy expressed in terms of exchange parameters based on the Heisenberg model, and the exchange parameters by self-consistently solving the resulting equations. To obtain $J_3$ along $b$ axis, two additional energy equations were established from ferromagnetic and antiferromagnetic $1 \times 2 \times 1$ supercells with spins aligned along the $b$ axis.

\begin{figure*}[htbp!]
    \centering
    \includegraphics[width=0.85\textwidth]{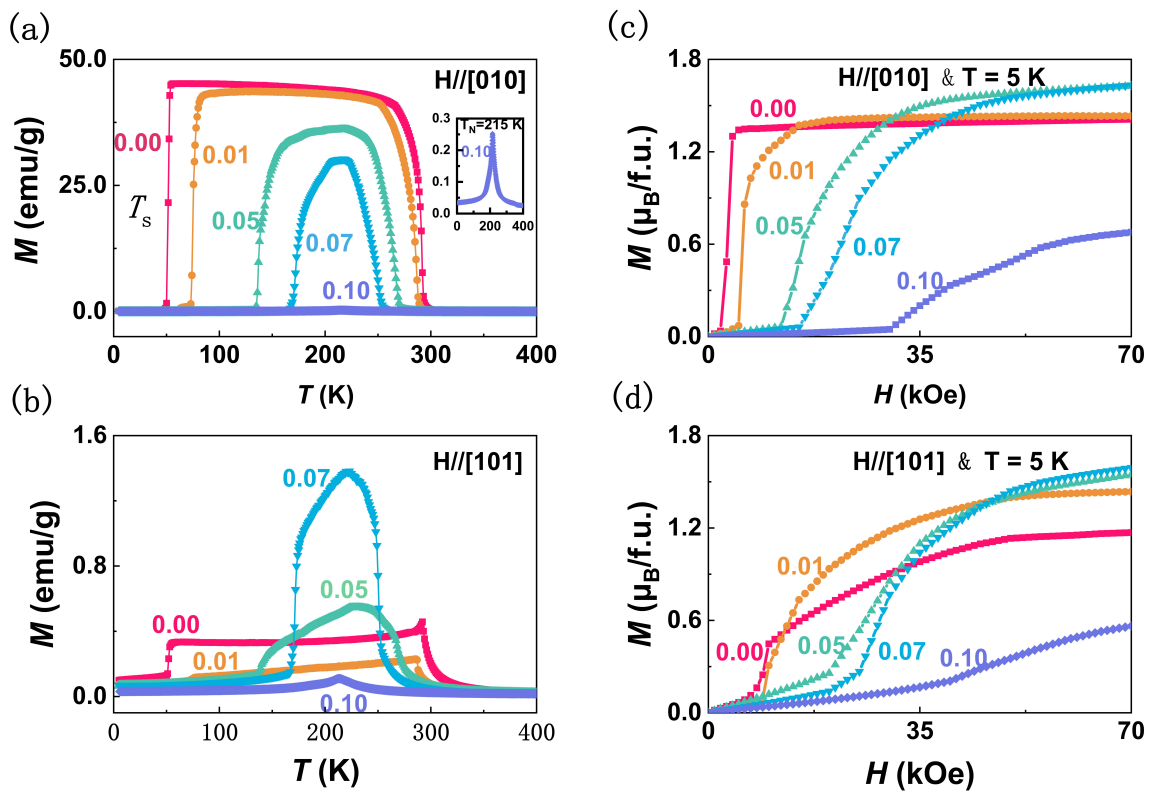}
    \vspace{0pt} 
    \caption{Magnetic properties of Mn$_{1-x}$Ru$_x$P single crystals. (a, b) Zero-field-cooled (ZFC) magnetization curves measured under an applied magnetic field of 100 Oe; (c, d) isothermal magnetization curves measured at 5 K. The data were acquired along the [010] crystallographic direction for panels (a) and (c), and along [101] for panels (b) and (d).}
    \vspace{0pt} 
    \label{Fig. 2}
\end{figure*}

First-principles calculations were performed using the VASP code with a plane-wave cutoff energy of 400 eV and a $6 \times 6 \times 1$ Monkhorst--Pack $k$-point mesh. Electronic self-consistency was converged to $10^{-6}$ eV per cell. To investigate the effects of lattice expansion on the helical phase, we constructed a $1 \times 1 \times 9$ supercell based on the experimental lattice parameters of Mn$_{1-x}$Ru$_x$P. In this model, Ru substitution was treated as an effective lattice dilation without explicitly incorporating Ru atoms. Magnetic exchange parameters were extracted by mapping the total energies of distinct spin configurations onto a Heisenberg Hamiltonian and solving the resulting system of linear equations\cite{34}. As illustrated in Fig.~1b, the interchain exchanges $J_1$, $J_1'$ and the intrachain exchange $J_2$ were determined using a $1 \times 1 \times 2$ supercell, involving ferromagnetic, interchain ferromagnetic, cross-antiferromagnetic, and intrachain antiferromagnetic magnetic configurations. Finally, the exchange parameter $J_3$ along the $b$-axis was derived from the energy difference between ferromagnetic and antiferromagnetic states in a $1 \times 2 \times 1$ supercell with spins aligned along the $b$-direction.

\section{Results and discussion}
\begin{figure*}[htbp!]
    \centering
    \includegraphics[width=0.85\textwidth]{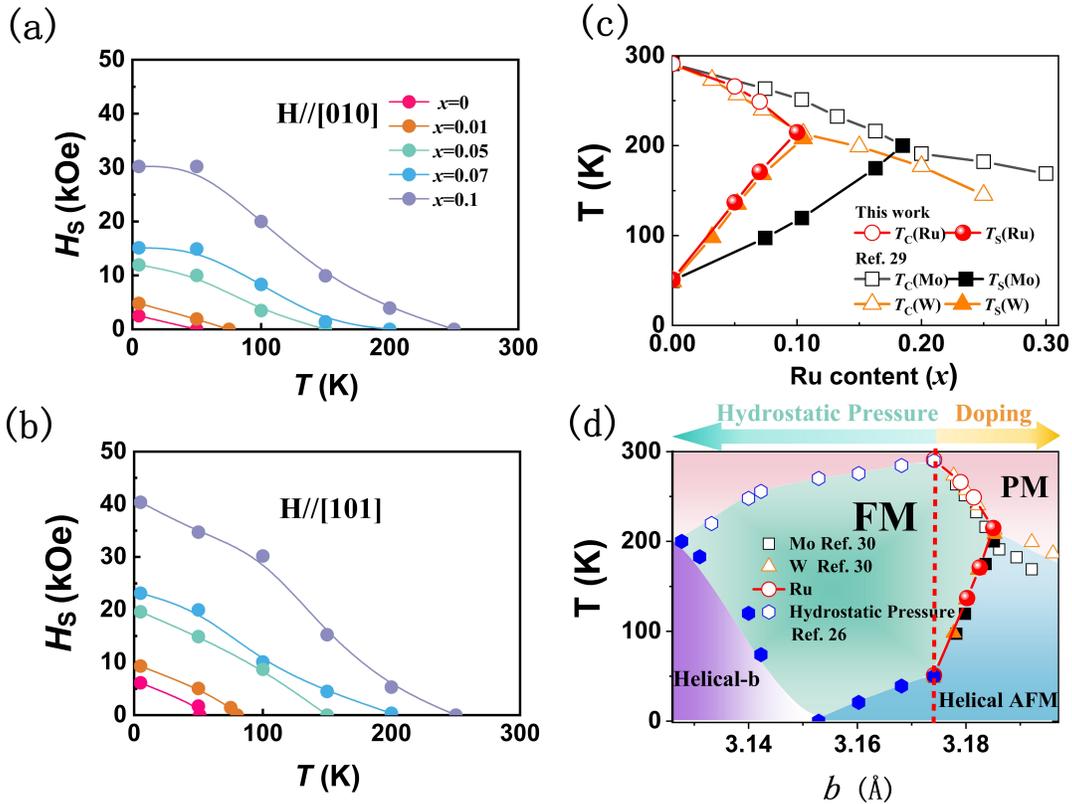}
    \vspace{0pt} 
    \caption{Scaling of magnetic phase transitions across chemically doped MnP.  (a) (b) Magnetic field--temperature ($H$--$T$) phase diagrams of Mn$_{1-x}$Ru$_x$P with the magnetic field applied along the [010] and [101] crystallographic directions, respectively.  (c) $T_{\rm C}$ and $T_{\rm S}$ as a function doping concentration $x$ in the Mn$_{1-x}$M$_x$P system (M = Ru, Mo, W) . (d) $T_{\rm S}$ and $T_{\rm C}$ plotted against the $b$-axis lattice parameter, integrating data from chemically doped MnP and hydrostatic pressure experiments on MnP.}
    \vspace{0pt} 
    \label{Fig. 2}
\end{figure*}

Figure 1(b) presents XRD patterns collected from the side facets of needle-like Mn$_{1-x}$Ru$_x$P single crystals. The well-defined (l0l) reflections confirm that the cross-sections correspond to the (101) planes and that crystal growth proceeds along the [010] direction\cite{6}. As $x$ increases, the (202) diffraction peak shifts monotonically from 46.14° to 45.82°, indicating lattice expansion induced by the larger ionic radius of Ru$^{2+}$ ($\sim 0.84\ \text{Å}$) compared to that of Mn$^{2+}$ ($\sim 0.67\ \text{Å}$). To evaluate phase purity and obtain precise lattice parameters, the crystals were ground into powders for powder XRD analysis. All patterns across the doping series index cleanly to the orthorhombic MnP-type structure with no detectable impurity phases, confirming single-phase formation and the successful substitution of Ru onto the Mn sites. Rietveld refinement, exemplified for the $x = 0.10$ crystal in Fig. 1(c), shows excellent agreement between the observed and calculated profiles. Although the refined lattice parameters in Fig. 1(d) exhibit a general linear dependence on $x$, their pronounced anisotropy underscores the directionally dependent response of the MnP lattice to Ru substitution. Specifically, at $x = 0.10$, the $a$- and $c$-axes expand by approximately $0.04$ Å, whereas the $b$-axis expands by only one-quarter of this magnitude.

\begin{figure}[htbp!]
    \centering
    \includegraphics[width=\linewidth]{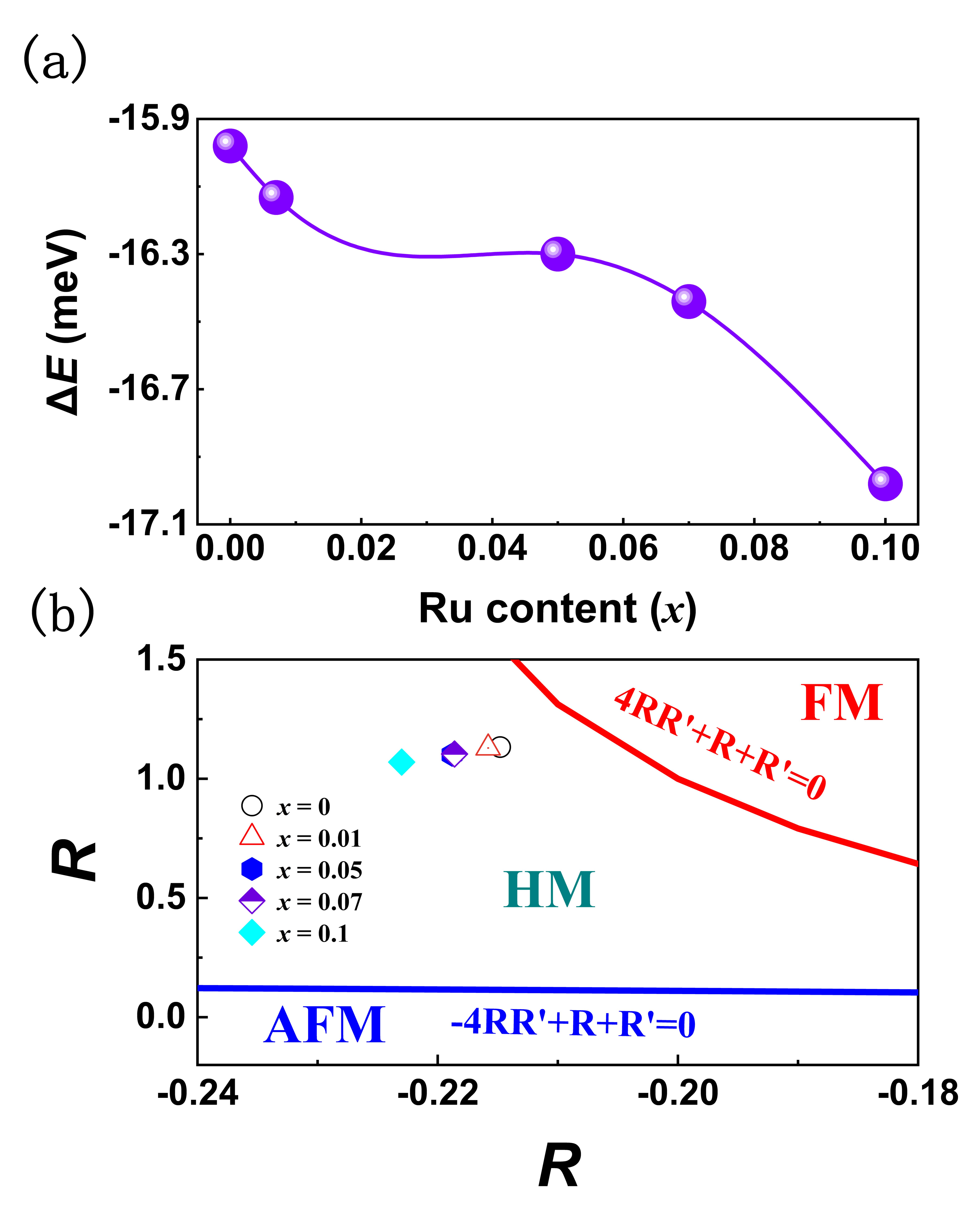}
    
    
    \vspace{6pt}
    
    \caption{Phase stability of the helical phase of MnP based on first-principles calculations. (a) Energy difference between the helical and ferromagnetic phases (left axis) as a function of Ru substitution level $x$. Calculations were performed using lattice parameters corresponding to different $x$ values, without explicitly substituting Mn atoms with Ru in the supercells. (b) Theoretical $R$--$R'$ phase diagram based on the Bertaut--Kallel model, where $R = J_1/J_2$ and $R' = J_1'/J_2$. The red and blue curves represent the boundaries separating the helical--ferromagnetic and helical--antiferromagnetic regions, respectively.}
    
    \vspace{6pt}
    \label{Fig.4}
\end{figure}

 Zero-field-cooling (ZFC) magnetization measurements along the [010] axis, presented in Fig. 2(a), reveal that pristine MnP undergoes a helical-to-ferromagnetic transition at 51 K upon warming, followed by a transition to the paramagnetic state at 291 K, which is consistent with the reported values\cite{9}. Ru substitution induces a pronounced increase in \(T_S\) accompanied by a gradual suppression of \(T_C\), thereby dramatically narrowing the temperature window of the ferromagnetic phase. By \(x = 0.10\), \(T_S\) and \(T_C\) merge at 215 K, signaling the complete suppression of the ferromagnetic phase. Measurements along the [101] direction, shown in Fig. 2(b), exhibit similar trends in transition temperatures but consistently lower magnetization than that measured along the [010] axis. In contrast to the monotonic decrease observed along the [010] axis, the magnetization along the [101] direction exhibits an anomalous enhancement at 200 K for \(x = 0.05\) and \(x = 0.07\), suggesting that Ru doping not only tunes transition temperatures but also modulates magnetic anisotropy of MnP.

The isothermal magnetization of Mn$_{1-x}$Ru$_x$P was measured at 5 K along the [010] and [101] directions and is shown in Figs. 2(c) and 2(d). The magnetic susceptibility within the low-temperature helical phase decreases systematically with increasing doping concentration $x$ along both directions, indicating a diminishing response of the helical order to external magnetic fields.For the undoped sample, a helical-to-ferromagnetic transition occurs at $H_S = 2.5$ kOe along the [010] axis, while along the [101] direction\cite{4}, it transforms into a fan-like structure via a helical-to-fan transition at $H_S = 6.7$ kOe. With Ru substitution, $H_S$ increases sharply in both directions, reaching 12.0 and 30.0 kOe along the [010] axis, and 20.0 and 40.0 kOe along the [101] direction for $x = 0.05$ and $x = 0.10$, respectively, indicating the progressive stabilization of the helical order against field-induced reorientation. The saturated magnetization at 70 kOe along both directions initially rises with doping, peaking at $1.6\ \mu_\text{B}$/f.u. for $x = 0.07$ before dropping sharply to $\sim 0.6\ \mu_\text{B}$/f.u. at $x = 0.10$. The initial increase likely arises from the larger local magnetic moment of Ru compared to Mn, whereas the subsequent suppression reflects a significant weakening of the ferromagnetic order and enhanced stabilization of the helical or fan magnetic phase under an external field.

Figs. 3(a) and 3(b) summarize the temperature dependence of the critical field of helical-to -ferromagnetic or-fan transition field ($H_S$) for the Mn$_{1-x}$Ru$_x$P system along the [010] and [101] directions, respectively. It is observed that both the critical temperature ($T_S$) and critical field ($H_S$) increase drastically with increasing $x$, demonstrating that Ru doping greatly enhances the stability of the helical phase. To investigate the mechanism underlying the stabilization of the helical order in Mn$_{1-x}$Ru$_x$P, we compared the doping dependence of $T_S$ and $T_C$ for Ru-doped MnP single crystals with those of W- and Mo-doped MnP polycrystals\cite{29}, as illustrated in Fig. 3(c). All three heavy-element-doped MnP systems exhibit a similar trend, in which $T_S$ increases and $T_C$ decreases with increasing doping concentration $x$, ultimately leading to the suppression of the ferromagnetic phase at the Lifshitz point\cite{31,32}. Nevertheless, Ru and W doping show nearly identical efficacy, fully suppressing ferromagnetism at doping level of 0.10, whereas Mo induces a markedly weaker shift, with $T_S$ and $T_C$ coinciding only at $x \approx 0.20$. Given the sensitivity of the Mn–Mn exchange interaction to lattice geometry\cite{22}, we investigated the evolution of $T_S$ and $T_C$ with lattice parameters in Ru-, W-, and Mo-doped MnP systems.

\begin{table*}[t]
    \centering
    \caption{Calculated nearest-neighbor Mn--Mn exchange interaction parameters, and the dimensionless ratios $R = J_1/J_2$ and $R' = J_1'/J_2$ based on the Bertaut--Kallel model.}
\label{tab:exchange_parameters}
    \label{Tab. 1}
    \setlength{\extrarowheight}{8pt}
    \setlength{\tabcolsep}{12pt} 
    \renewcommand{\arraystretch}{1} 
    \begin{tabular}{cccccccc}
       \hline
$x$ & $J_2$ (meV) & $J_1$ (meV) & $J_1'$ (meV) & $J_3$ (meV) & $R = J_1/J_2$ & $R' = J_1'/J_2$ \\
\hline
0.00 & 32.70 & 37.04 & -7.03 & 16.38 & 1.329 & -0.215 \\
0.01 & 32.30 & 36.48 & -6.97 & 16.30 & 1.129 & -0.216 \\
0.05 & 32.68 & 36.10 & -7.15 & 16.11 & 1.104 & -0.219 \\
0.07 & 32.00 & 35.23 & -7.00 & 15.90 & 1.101 & -0.219 \\
0.10 & 31.40 & 33.60 & -6.99 & 15.65 & 1.070 & -0.223 \\
\hline
    \end{tabular}
\end{table*}

 Across all three systems, the lattice constants $a$ and $c$ vary identically with the doping level $x$, whereas the change in $b$ for Mn$_{1-x}$Mo$_x$P is only half that observed in Mn$_{1-x}$Ru$_x$P and Mn$_{1-x}$W$_x$P\cite{30}. Remarkably, when plotted against the b-axis length on the right side of Fig. 3(d), both $T_S$ and $T_C$ data collapse onto two separate straight lines, revealing a direct correlation between lattice distortion and magnetic transitions in MnP. This scaling behavior initially suggests that, regardless of the specific dopant, lattice distortion, particularly along the $b$-direction, governs the evolution of the magnetic phase. Qualitative analysis reveals steep sensitivities of $\frac{dT_C}{db} = 0.69 \times 10^4\ \text{K}/\text{\AA}$ and $\frac{dT_S}{db} = -1.59 \times 10^4\ \text{K}/\text{\AA}$ along the $b$-axis, whereas the responses to variations in $a$ and $c$ are negligible, with coefficients of $\frac{dT_C}{da} \approx \frac{dT_C}{dc} \approx 0.037 \times 10^3\ \text{K}/\text{\AA}$ and $\frac{dT_S}{da} \approx \frac{dT_S}{dc} \approx -0.033 \times 10^4\ \text{K}/\text{\AA}$ for $T_C$ and $T_S$, respectively.

However, a critical discrepancy emerges when comparing chemical doping with hydrostatic pressure. While hydrostatic pressure also induces a dominant linear contraction of the $b$-axis with minimal changes in $a$ and $c$\cite{23}, the resulting magnetic responses differ fundamentally on the left side of Fig. 3(d). Under pressure, $T_C$ decreases monotonically, whereas $T_S$ exhibits a non-monotonic upturn below $b \approx 3.15$ Å.
Notably, the derived $\frac{dT_S}{db}$ and $\frac{dT_C}{db}$ magnitudes under hydrostatic pressure are only one-third and one-fourth of those observed under chemical doping, a discrepancy that warrants further investigation. This dichotomy is further corroborated by uniaxial pressure measurements along the b-axis, which reports a large negative pressure coefficient for $T_S$ ($-46.5\ \text{K·GPa}^{-1}$) and a positive coefficient for $T_C$ ($+38.9\ \text{K·GPa}^{-1}$) \cite{22,33}. Together, these results demonstrate that, despite differing physical origins, chemical doping, hydrostatic pressure, and uniaxial stress all correlated with the changes in Mn-Mn distances—especially along the b-direction.

To isolate the role of lattice expansion in stabilizing the helical phase of MnP, we performed first-principles calculations using experimentally refined lattice parameters, explicitly excluding Ru atoms to focus solely on structural effects. The relative stability of the helical and ferromagnetic phases was quantified by the energy difference $\Delta E = E_{\text{HM}} - E_{\text{FM}}$\cite{21}. As shown in Fig. 4(a), $\Delta E$ decreases from $-15.9\ \text{meV}$ to $-17.0\ \text{meV}$ as $x$ increases from 0.00 to 0.10, confirming that lattice expansion enhances the thermodynamic stability of the helical ground state. This trend aligns with the experimental rise of both $T_S$ and $H_S$ in Figs. 2(a) and 2(b). To elucidate the microscopic origin of this stabilization, we computed the evolution of the nearest-neighbor Mn–Mn exchange coupling parameters as a function of $x$. As listed in Table 1, $J_1$, $J_2$, and $J_3$ are positive, while $J_1'$ is negative—a hierarchy consistent with the reported signs of the exchange parameters for helimagnetism in MnP \cite{34}. With increasing $x$ from 0.00 to 0.10, $J_1$, $J_2$, and $J_3$ decrease by 3.44, 1.30 and 0.73 meV, respectively, while $J_1'$ remains nearly a constant around $-7.00\ \text{meV}$. This selective suppression of ferromagnetic exchange interactions strengthens the relative weight of antiferromagnetic interactions, thereby reinforcing the helical magnetic order. The concurrent reduction in $J_1$, $J_2$, and $J_3$ also explains the experimentally observed monotonic decrease in $T_C$ with doping.

Bertaut and Kallel described the magnetic phase diagram of the MnP-type helical structure by introducing dimensionless ratios $R = J_1/J_2$ and $R' = J_1'/J_2$, where the ferromagnetic-helical and antiferromagnetic-helical phase boundaries are given by the hyperbolas $4RR' + R + R' = 0$ and $-4RR' + R + R' = 0$, respectively\cite{35,36}. Using the computed exchange coupling constants, we derived $R$ and $R'$ (listed in the last two columns of Table 1) and plotted the corresponding $(R, R')$ points in the phase diagram of Fig. 4(b). Undoped MnP lies within the helical region but close to the ferromagnetic-helical boundary—consistent with the sensitivity of its helical order to pressure and chemical doping. As $x$ increases from 0.00 to 0.10, the systematic reduction in both $R$ and $R'$ shifts the points progressively deeper into the helical phase, indicating enhanced thermodynamic stability of the helical state driven by lattice expansion. This trend is quantitatively consistent with the concurrent rise in $T_S$ from $51\ \text{K}$ to $215\ \text{K}$ and $H_S$ from $2.5\ \text{kOe}$ to $30\ \text{kOe}$ measured along the [010] direction at 5 K. Together, these results demonstrate that lattice expansion, induced by substituting Mn with heavy elements such as Ru, Mo, or W, modulates competing magnetic exchange interactions to systematically enhance the stability of the helical phase in MnP.

\section{Conclusion}
We demonstrate that Ru substitution in Mn$_{1-\textit{x}}$Ru$_\textit{x}$P ($0 \le x \le 0.10$) single crystals induces a highly anisotropic lattice expansion that dramatically stabilizes the helical phase. This structural modulation elevates the $T_{\mathrm{\textit{S}}}$ from 51 K to 215 K and increases the critical field along the [010] direction at 5 K from 2.3 kOe to 30.0 kOe. A unified analysis encompassing Ru-, Mo-, and W-doped MnP reveals a universal linear scaling of both $T_{\mathrm{\textit{S}}}$ and $T_{\mathrm{\textit{C}}}$ with the $\textit{b}$-axis lattice parameter, characterized by coefficients of $\frac{dT_\textit{S}}{d\textit{b}}=1.59 \times 10^4\ \text{K·Å}^{-1}$ and $\frac{dT_\textit{C}}{d\textit{b}}=0.69 \times 10^4\ \text{K·Å}^{-1}$  ---values significantly exceeding those associated with the $\textit{a}$- or $\textit{c}$-axes. This dominant role of the $\textit{b}$-axis as the primary structural lever governing helical stability is consistent with the hydrostatic and uniaxial pressure studies. First-principles calculations indicate that the anisotropic expansion selectively attenuates ferromagnetic exchange while preserving antiferromagnetic coupling between nearest-neighbor Mn atoms, thereby intensifying magnetic frustration to lock in the helimagnetic state. Collectively, these findings establish chemical pressure via targeted elemental substitution as a robust and generalizable paradigm for engineering the helical magnetic ground state in MnP, paving the way for its practical deployment in chiral spintronic devices.

\begin{acknowledgments}
The authors gratefully acknowledge the financial support provided by the National Natural Science Foundation of China (Grant No.~51971087), the ``333 Talent Project'' of Hebei Province (Grant No.~C20231105), the Basic Research Project of Shijiazhuang Municipal Universities in Hebei Province (Grant No.~241790617A), the Central Guidance on Local Science and Technology Development Fund of Hebei Province (Grant No.~236Z7606G), and the Science Foundation of Hebei Normal University (Grant No.~L2024B08). These funding sources have been instrumental in facilitating the completion of this research.
\end{acknowledgments}

\end{document}